\documentclass[referee]{raa}            

\usepackage{graphicx,times}             
\usepackage{natbib}
\usepackage{amssymb,amsmath}

\usepackage[a4paper=true,dvipdfm=true,pagebackref=true]{hyperref}
\hypersetup{colorlinks = true, linkcolor = green, anchorcolor = red, citecolor = blue, filecolor = red, pagecolor = red, urlcolor = red}

\begin{document}

\title{Automatic Detection and Correction Algorithms for Magnetic Saturation in the SMFT/HSOS longitudinal  Magnetograms}

   \volnopage{Vol.0 (20xx) No.0, 000--000}      
   \setcounter{page}{1}          

   \author{*Hai-qing, Xu \inst{1}, Suo Liu \inst{1}, *Jiang-tao Su \inst{1,2}, Yuan-yong Deng \inst{1,2}, Andrei Plotnikov \inst{3}, Xian-yong Bai \inst{1,2}, Jie Chen \inst{1}, Xiao Yang \inst{1}, Jing-jing Guo \inst{1}, Xiao-fan Wang \inst{1} and Yong-liang Song \inst{1}}
\institute{Key Laboratory of Solar Activity, National Astronomical Observatories, Chinese
Academy of Sciences, Beijing, 100101, China; {\it xhq@bao.ac.cn; sjt@bao.ac.cn}\\
          \and
             School of Astronomy and Space Sciences, University of Chinese Academy of Sciences,
19 A Yuquan Road, Shijingshan District, Beijing 100049, China\\
         \and
          Crimean Astrophysical Observatory, Russian Academy of Sciences, Nauchny,
Crimea, 298409, Russia\\
\vs\no
   {\small Received~~20xx month day; accepted~~20xx~~month day}}

\abstract{longitudinal magnetic field often suffers the saturation effect in strong magnetic field region when the measurement performs in a single-wavelength point and linear calibration is adopted. In this study, we develop a method \textbf{that} can judge the threshold of saturation in Stokes $V/I$ observed by the Solar Magnetic Field Telescope (SMFT) and \textbf{correct for} it automatically. \textbf{The procedure is that} first perform the second-order polynomial fit to the Stokes $V/I$ \textit{vs} $I/I_{m}$ ($I_{m}$ is the maximum value of Stokes $I$) curve to estimate the threshold of saturation, then reconstruct Stokes $V/I$ in strong field region to \textbf{correct for} saturation. The algorithm is proved to be effective by comparing with the magnetograms obtained by the Helioseismic and Magnetic Imager (HMI). The accurate rate of detection and correction  for saturation is $\sim$99.4\% and $\sim$88\% respectively among 175 active regions.  The advantages and disadvantages of the algorithm are discussed.
}

   \authorrunning{H.-Q. Xu et al.}            
   \titlerunning{Detection and Correction for Magnetic Saturation }  

   \maketitle

%

\section{Introduction}           
\label{sect:intro}

The study of solar magnetic field has always been a core topic in solar physics. Some major unsolved scientific problems in the study of solar physics, such as the generation of solar cycle, coronal heating, the origin of solar eruptions and so on, are all related to the solar magnetic field. The magnetic field of sunspots was first investigated by \cite{Hale+1908}. It is known that \textbf{generally,} the present magnetographs measure solar polarized light (present as Stokes parameters $I$, $Q$, $U$, and $V$) rather than magnetic fields. Under a certain atmospheric models and assumptions, the solar magnetic field is obtained \textbf{through} inversion according to the radiation transfer theory. The solar magnetic field has been observed for one century and many interesting results have been presented from these observations. \textbf{Yet,}  there are still some basic questions on the measurements of solar magnetic fields \textbf{waiting} to be carefully analyzed (\citealt{Zhang+2019}). \cite{Svalgaard+etal+1978} found the magnetograph was saturated when magnetic field is very strong. \cite{Ulrich+etal+2002} discussed reasons and treatment of saturation effects in the Mount Wilson 150 foot tower telescope system in detail. They pointed out that most spectral lines used for magnetic measurements are subject to this saturation effect for at least some parts of their profile. \cite{Liu+etal+2007} found another type of saturation in sunspot umbrae observed by the Michelson Doppler Imager on the Solar and Heliospheric Observatory (MDI/SOHO) caused by the 15-bit on-board numerical treatment used in deriving the MDI magnetograms. The saturation effect can be eliminated by using the information on spectral line, e.g., the Spectro-polarimeter (SP) onboard
Hinode (\citealt{Kosugi+etal+2007}) obtains spectral profile with a wide spectral range, and the Helioseismic and Magnetic Imager on board the Solar Dynamics Observatory (HMI/SDO, \citealt{Schou+etal+2012}) obtains spectral profile with six wavelength points. The saturation effect needs to be \textbf{corrected} by some supplementary methods if the polarized light is measured in a single-wavelength point and linear calibration is adopted. \cite{Chae+etal+2007} performed cross-calibration of Narrow-band Filter Imager (NFI) Stokes $V/I$ and longitudinal magnetic field acquired by the SP, and proposed to use two different linear relationships of longitudinal magnetic field and Stokes $V/I$ from Hinode/NFI to \textbf{correct for saturation}. \cite{Moon+etal+2007} used a pair of MDI intensity and magnetogram data simultaneously observed, and the relationship from the cross-comparison between the SP and MDI flux densities to \textbf{correct for saturation} in magnetic field obtained by MDI. \cite{Guo+etal+2020} explored a nonlinear calibration  method to deal with the saturation problem, which used a multilayer perceptron network.

The Solar Magnetic Field Telescope (SMFT) at the Huairou Solar Observing Station (HSOS) of the National Astronomical Observatories of China is a 35 cm vacuum telescope equipped with a birefringent
filter for wavelength selection and KD*P crystals to modulate polarization signals. The Fe {\sc i} 5324.19 {\AA} line is used for measurements. A vector magnetogram is built using four narrow-band (0.125 {\AA}) Stokes $I, Q, U$ and $V$ maps. The center wavelength of the filter can be tuned and
is normally at $-0.075$ {\AA} for the measurements of longitudinal magnetic fields and at the line center for the transversal magnetic fields (\citealt{Ai+Hu+1986}). It has been observing vector
magnetic fields for more than 30 years. The theoretical calibration for SMFT vector magnetogram was first made by \cite{Ai+etal+1982}. Several different methods of the magnetic field calibration under the weak-field assumption have been done since then. \cite{Wang+etal+1996a} used an empirical calibration and a velocity calibration methods to calibrate the longitudinal magnetograms. \cite{Su+Zhang+2004} used 31 points of the Fe {\sc i} 5324.19 {\AA} spectral line profile to derive vector magnetic field by non-linear least squares fitting technique. \cite{Bai+etal+2014} improved the calibration process by fitting the observed full Stokes information using six points \textbf{of}  the profile of Fe I 5324.19 {\AA} line, and the analytical Stokes profiles under the Milne-Eddington atmosphere model, adopting the Levenberg-Marquardt least-squares fitting algorithm. However, the routine measurements of Stokes $I, Q, U$ and $V$ parameters by SMFT are being performed in a single-wavelength point. The longitudinal magnetic field is reconstructed by equation \ref{eq0}:
\begin{equation}  \label{eq0}
 B_{L}^{SMFT} =C_{L}{V \over I}
 \end{equation}
where $C_{L}$ is the calibration coefficient inferred from the aforementioned calibration methods. This linear calibration will result in the saturation when magnetic field is strong. \cite{Plotnikov+etal+2019} made an attempt to improve the routine magnetic field measurements of SMFT by introducing non-linear relationship between the Stokes $V/I$ and longitudinal magnetic field. They performed cross-calibration of SMFT data and magnetograms provided by HMI to determine the form of the relationship. They found that the magnetic field
saturation inside sunspot umbra can be eliminated by using non-linear relationship between Stokes $V/I$ and longitudinal magnetic field. They also discussed the influence of saturation effect \textbf{on solving the 180 degree ambiguity} of the transversal magnetic field. They manually chose the threshold for separating pixels into
two subsets of strong and weak magnetic field, which is not convenient for dealing with large data sample.

In this paper, we attempt to develop a method which can judge the threshold of saturation in SMFT longitudinal field and \textbf{correct for} saturation automatically. One purpose of this study is to \textbf{correct for saturation} in longitudinal magnetic field obtained by SMFT since 1987. Another purpose is to prepare calibration technique for the Full-disk vector MagnetoGraph (FMG, \citealt{Deng+etal+2019}) which is one payload \textbf{onboard} the Advanced Space-based Solar Observatory (ASO-S, \citealt{Gan+etal+2019}) that will be launched in early 2022. The routine observations for the FMG  will be taken at one wavelength position of the Fe I 5324.179 \AA{} (\citealt{Su+etal+2019}). The magnetic field will be suffered saturation effect if the linear calibration is adopted.

\section{Observations}

The raw data registered by SMFT \textbf{are} left and right polarized light. The Stokes $V/I$ and $I$  are calculated as following:
\begin{equation}  \label{eq1}
\begin{aligned}
& {{V} \over {I}} ={{V_{l}-V_{r}}\over {V_{l}+V_{r}}} \\
& I=V_{l}+V_{r} \\
\end{aligned}
\end{equation}
where $V_{l}={{I+V} \over {2}}$ and $V_{r}={{I-V} \over {2}}$ represent modulated filtergrams. After this process, the influence of flat field is eliminated. The pixel size of SMFT data is approximately $0.29''\times0.29''$ since 2012 and the \textbf{spatial}  resolution is approximately $2''$ produced by local seeing effect. We selected 9 active regions (AR) in between 2013 and 2015 for case study and 175 ARs in 2013 for statistical study. The data were performed \textbf{by} $4 \times 4$ pixels median filtering to reduce the noise.

To check the effectiveness of correction method for saturation, we downloaded the co-temporal magnetograms of the selected 9 ARs from HMI/SDO. HMI is a full-disk filtergraph that measures the profile of photospheric Fe~I 6173~\AA\ line at six wavelength positions in two polarization states to derive the longitudinal magnetic field. The spatial resolution of the instrument is $1''$ with $0.5''\times0.5''$ pixel size. In order to perform a detailed pixel by pixel comparison, \textbf{HMI magnetograms} were rotated for the $p$-angle correction and reduced \textbf{in} spatial resolution to $2''$ by \textbf{a} 2-D Gauss-smooth function. Both data are re-scaled to the pixel size of $0.5''\times0.5''$. Then the same region \textbf{that} includes the maximized size of sunspots are selected and shifted each other to determine the optimal registration.

\section{Method}

\begin{figure}
   \centerline{\includegraphics[width=14cm,clip=]{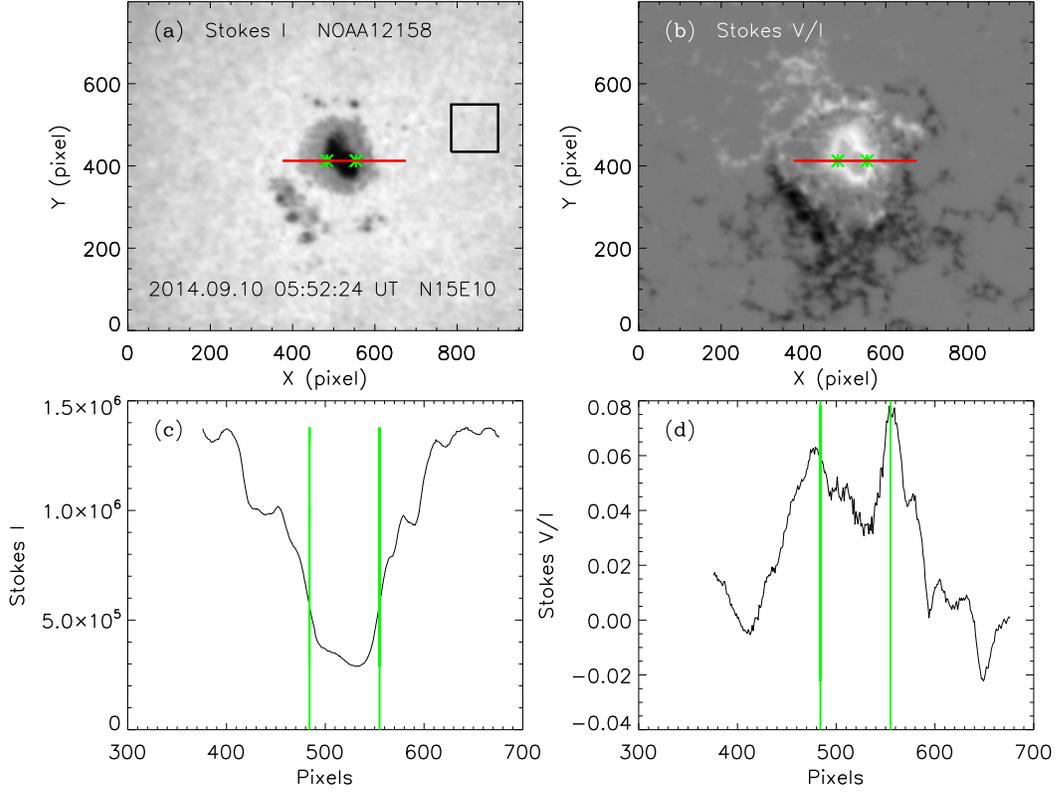}}
   \caption{Panels (a) and (b) are maps of Stokes $I$ and $V/I$ observed on 2014 September 10 by SMFT.
The rectangle region in panel (a) is used to calculate $I_{c}$. Panels (c) and (d) show the distribution of Stokes $I$ and $V/I$ along the red line; The green lines indicate the saturation locations marked by asterisks.} \label{spot}
\end{figure}

The studies show that there is a relationship between the continuum intensity and magnetic field, and the smallest intensity always corresponding to the largest magnetic field (\textit{e.g.}, \citealt{Martinez+Vazquez+1993, Norton+Gilman+2004, Leonard+Choudhary+2008} ). Figure \ref{spot}(a) and (b) show the maps of Stokes $I$ and $V/I$ for active region NOAA12158 observed on 2014 September 10 by SMFT. Figure \ref{spot}(c) and (d) show the distribution of Stokes $I$ and $V/I$ along the red line. The Stokes $I$ decreases to the minimum in the sunspot center, but the Stokes $V/I$ stops increasing at the points marked by green lines (asterisks) corresponding to sunspot umbrae. This phenomenon is called magnetic saturation. If \textbf{performing} linear calibration, the longitudinal magnetic field will \textbf{get weakened} in sunspot umbrae \textbf{in comparing with its surrounding area}.

Next, we will show two examples to give a detail description for the detection and correction method for saturation effect.

\subsection{Detection Algorithm for Magnetic Saturation}

\begin{figure}
   \centerline{\includegraphics[width=14cm,clip=]{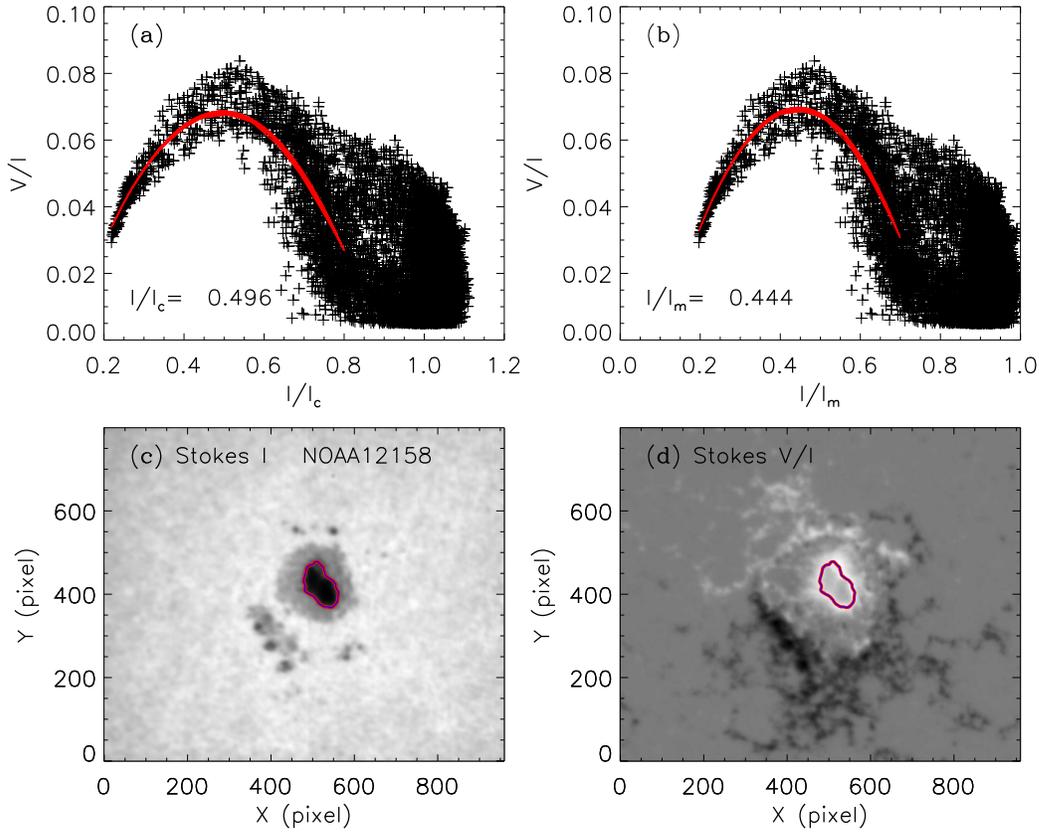}}
   \caption{Scatter plots of $|V/I|$ \textit{vs} $I/I_{c}$ (panel (a)) and $I/I_{m}$ (panel (b)) for NOAA12158. The red line is the best-fit second-order polynomial and $I/I_{c}$ ($I/I_{m}$) marked in panel is corresponding to the apex. The red and blue contours in panels (c) and (d) represent $I/I_{c}$=0.496 and $I/I_{m}$=0.444. $V/I$ uses the absolute values in the plots,  similarly hereinafter.} \label{thedfit1}
\end{figure}

The relationship between $|V/I|$ and $I/I_{c}$ ($I/I_{m}$) for NOAA12158 is given in Figure \ref{thedfit1}(a) ((b)). $I_{c}$  is the median value of Stokes $I$ within the rectangle region in Figure \ref{spot}(a). $I_{m}$ is the maximum value of Stokes $I$ within the whole active region. $|V/I| $ first increases with Stokes $I$ decreasing (going to sunspot center), then decreases. It is found that the second-order polynomial (red line) gives a well fit when $|V/I|$ $>$ 0.02 and $I/I_{c}$ $\leq$ 0.8 ($I/I_{m}$ $\leq$ 0.7). \cite{Moon+etal+2007} found similar relationship between MDI flux density and intensity \textbf{in magnetic saturation regions}, and used the second-order polynomial to separate the strong and weak field area. The pixels are separated into two parts by the apex. It is easy to calculate coordinates of the apex by fitting coefficient. We find that the corresponding $I/I_{c}$ is 0.496 and $I/I_{m}$ is 0.444, which are denoted in Figure \ref{thedfit1}(c) and (d) by blue and red contours. Although the value of $I/I_{c}$ and $I/I_{m}$ is different, the region in Stokes $I$ and $|V/I|$ maps is the same. When $I/I_{c}$ $<$ $0.496$ ($I/I_{m}$ $<$ 0.444), it is corresponding to the sunspot umbrae where the $|V/I|$ suffered from saturation. So we may use this value as the threshold to detect \textbf{the} saturation \textbf{regions} in longitudinal magnetic field.

\begin{figure}
   \centerline{\includegraphics[width=14cm,clip=]{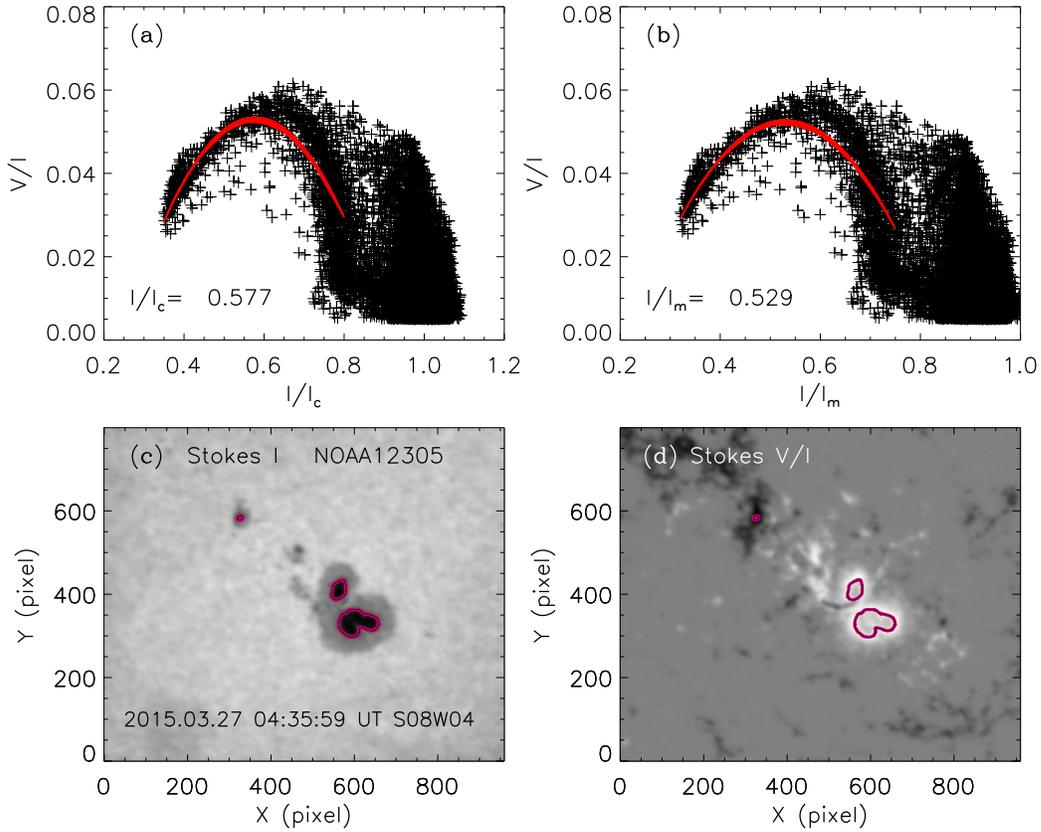}}
   \caption{Similar as Figure \ref{thedfit1}, but for NOAA12305.} \label{thedfit2}
\end{figure}

We performed the same analysis for NOAA12305 which includes multiple sunspots. We used the observation on 2015 Mar 27. The similar relationship was found between $|V/I|$ and $I/I_{c}$ ($I/I_{m}$) as NOAA12158 when $|V/I|$ $>$ 0.02 and $I/I_{c}$ $\leq$ 0.8 ($I/I_{m}$ $\leq$ 0.75). The $I/I_{c}$ ($I/I_{m}$) is 0.577 (0.529) corresponding to the apex of the second-order polynomial (red line in Figure \ref{thedfit2}(a) and (b)), which are denoted in Figure \ref{thedfit2}(c) and (d) by blue and red contours. The region of saturation can also be detected accurately for multiple sunspots.

We listed the thresholds for detecting saturation obtained by the above method for 9 active regions in Table \ref{tab1}. It can be seen that the thresholds are different for each active region. So it is necessary to calculate the threshold for individual active region. There is no difference in detecting the saturation region using $I/I_{c}$ and $I/I_{m}$, but $I/I_{m}$ has more advantages than $I/I_{c}$ \textbf{in} automatic detection.

\begin{table}
\bc
\caption[]{Threshold \textbf{of} detecting magnetic saturation for 9 ARs observed by SMFT\label{tab1}}
 \begin{tabular}{ccccccccccc}
  \hline\noalign{\smallskip}
NOAA& Date& Position&$I/I_{c}$&$I/I_{m}$\\
  \hline\noalign{\smallskip}
11658&2013.11.19&S11W10&0.602&0.497\\
11899&2013.11.16&N06W03&0.420&0.382\\
11960&2014.01.25&S14E01&0.661&0.585\\
12027&2014.06.07&N13W01&0.579&0.492\\
12055&2014.05.12&N10W02&0.664&0.557\\
12149&2014.08.27&N10E11&0.543&0.472\\
12158&2014.09.10&N15E10&0.496&0.444\\
12305&2015.03.27&S08W04&0.577&0.529\\
12325&2015.04.19&N04E02&0.638&0.582\\
  \noalign{\smallskip}\hline
\end{tabular}
\ec
\end{table}

\subsection{Correction Algorithm for Magnetic Saturation}

\begin{figure} 
\centering \resizebox{0.98\textwidth}{!}
{\includegraphics[]{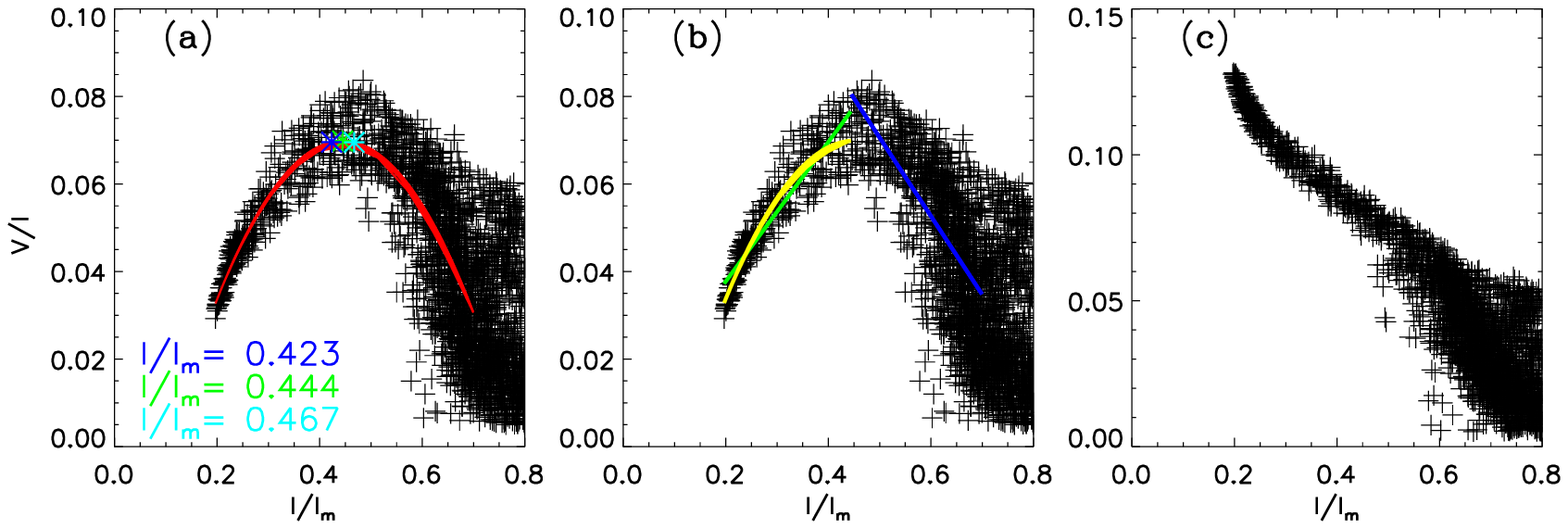}}
\resizebox{0.98\textwidth}{!}
{\includegraphics[]{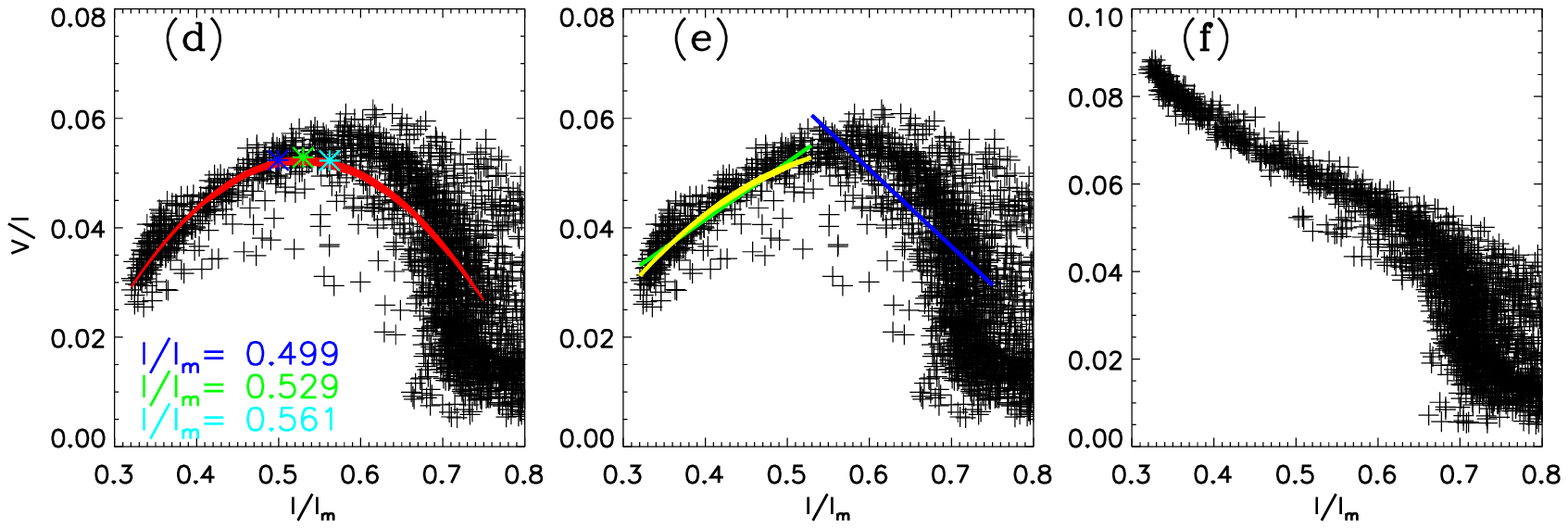}}
\caption{%
Scatter plots of $|V/I|$ \textit{vs} $I/I_{m}$ for NOAA12158 (panels (a)--(c)) and NOAA12305 (panels (d)--(f)). The green asterisks in panels (a) and (d) present the apex of the second-order polynomial fit (red line). The cyan and blue asterisks present the $\pm 1\sigma$ uncertainty of apex. The corresponding $I/I_{m}$ values are marked using the same color as \textbf{those} asterisks. The green and blue \textbf{(yellow)} lines are the linear \textbf{(the second-order polynomial)} fit to the data in panels (b) and (e). Panels (c) and (f) show the scatter plots of $|V/I|$ \textit{vs} $I/I_{m}$ after correcting magnetic saturation.}\label{scatter}
\end{figure}

\begin{figure} 
 \centering \resizebox{0.98\textwidth}{!}
{\includegraphics[]{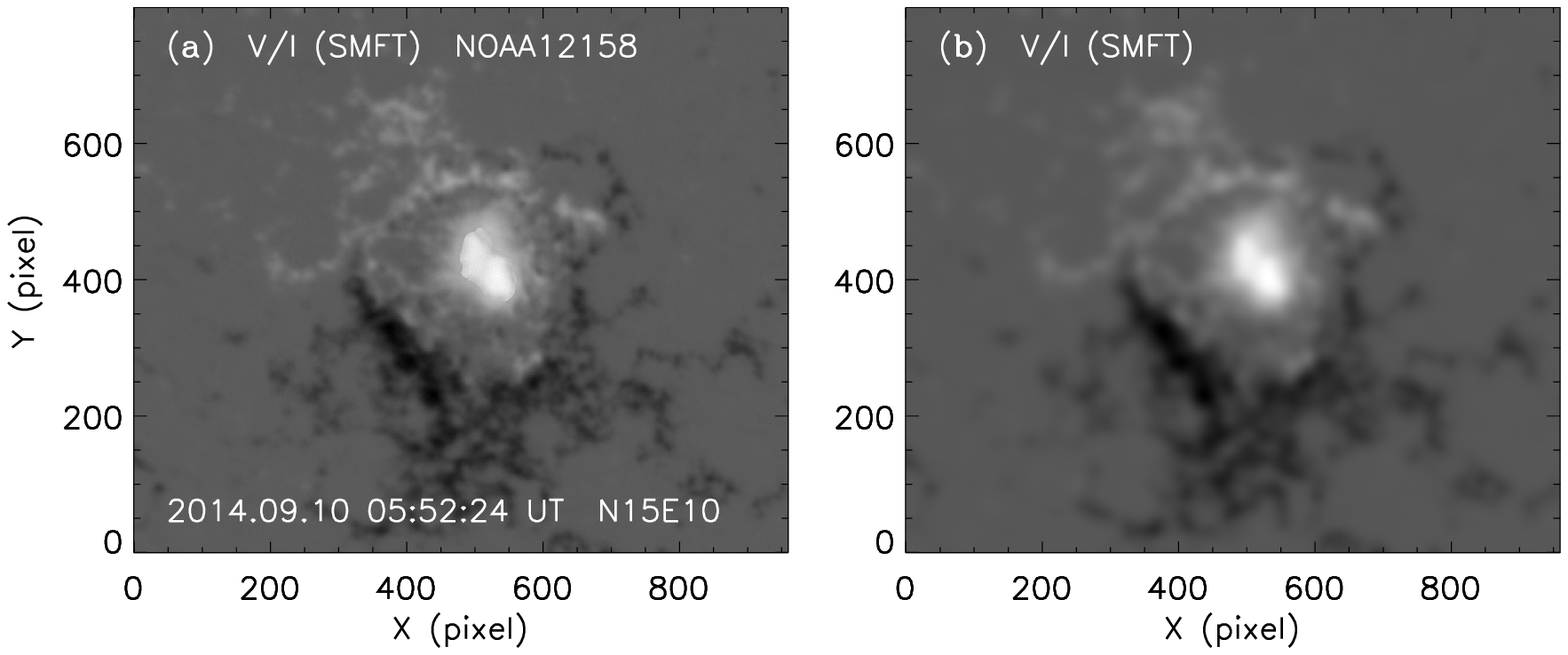}}
\resizebox{0.98\textwidth}{!}
{\includegraphics[]{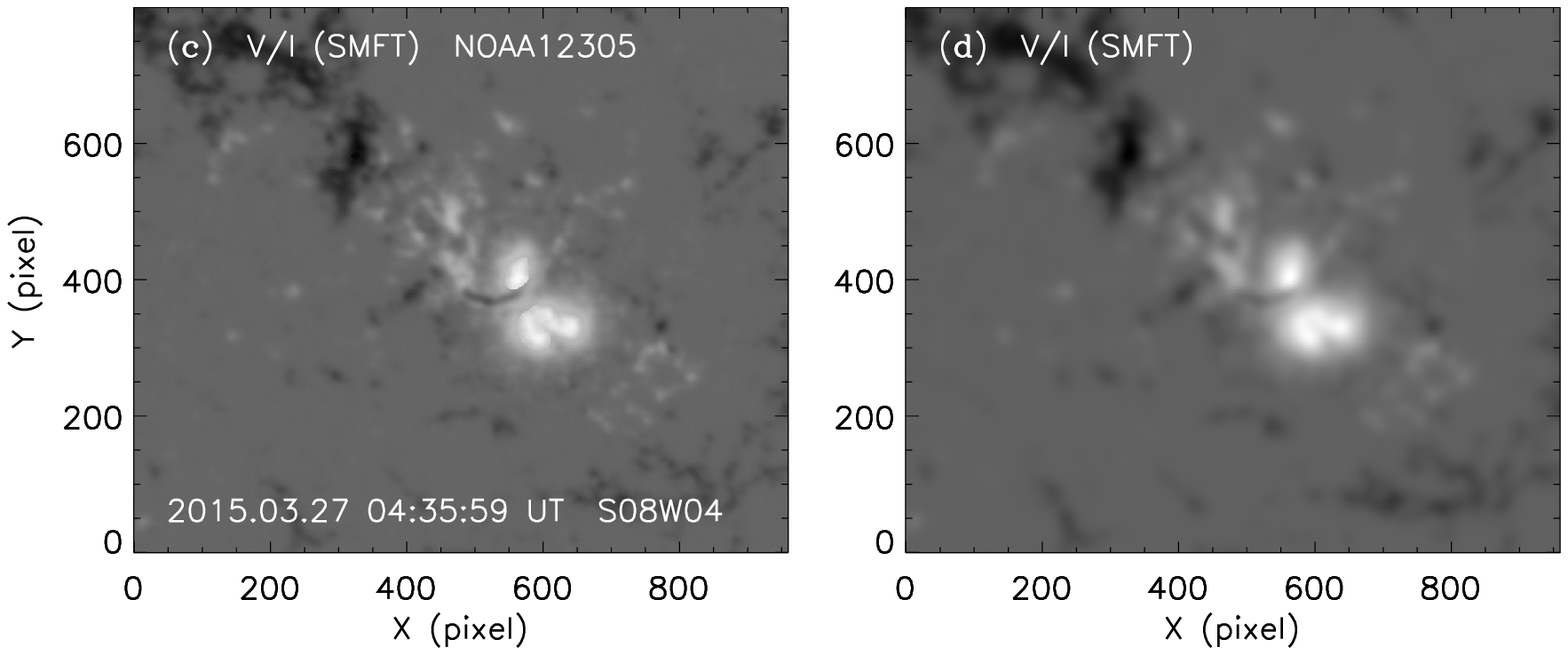}}
\caption{Maps of Stokes $V/I$ after correcting magnetic saturation for NOAA12158 (panels (a) and (b)) and NOAA12305 (panels (c) and (d)). The data in  panels (b) and (d) are precessed with  Gauss-smooth.}\label{cmap}
\end{figure}

We take the above two active regions as examples to show how \textbf{to} correct saturation effect. We re-plot  $|V/I|$ \textit{vs} $I/I_{m}$ in Figure \ref{scatter}. The correction procedure for magnetic saturation is as following:\\
(1) The threshold $I_{0}$ for occurrence of magnetic saturation is determined by the above algorithm, which corresponds to green asterisks in Figure \ref{scatter}(a) and (d).\\
(2) The pixels are separated into two parts by threshold $I_{0}$. \textbf{Those with} $I/I_{m}$ $<$ $I_{0}$ are suffered from saturation effect. \textbf{Both linear (green lines in Figure \ref{scatter}(b) and (e)) and the second-order polynomial functions (yellow lines in Figure \ref{scatter}(b) and (e)) are used to fit the scatter plots for saturation data. The green and yellow lines almost overlap. For its simplicity, we finally choose the linear functions to fit the scatter plots for both saturation and good data.} The fitting coefficients are ($a_{1}$, $c_{1}$) and ($a_{2}$, $c_{2}$) corresponding to green and blue lines in Figure \ref{scatter}(b) and (e), respectively. $a_{1}$ and $a_{2}$ are slopes, $c_{1}$ and $c_{2}$ are constants. Using equation \ref{eq2} to calculate Stokes $V/I$ for pixels where $I/I_{m}$ $<$ $I_{0}$,\\
\begin{equation}  \label{eq2}
{V \over I}^s =\left|{a_{2} \over a_{1}}\left(\left|{V \over I}\right|-c_{1}\right)+ c_{2}\right|\cdot sign\left({V \over I}\right)\\.
\end{equation}
(3) After re-calculating, the Stokes $V/I$ maps are shown in Figure \ref{cmap}(a) and (c). It can be seen that the saturation in sunspot umbrae has been eliminated. But the discontinuity at the boundary of umbrae and penumbrae can be seen. To eliminate this discontinuity, we calculate the $\pm 1\sigma$ uncertainty of $I_{0}$ corresponding to the cyan ($I_{+\sigma}$) and blue ($I_{-\sigma}$) asterisks in Figure \ref{scatter}(a) and (d). For pixels where $I_{-\sigma}$ $<$ $I/I_{m}$ $<$ $I_{+\sigma}$, $V/I$ is calculated by interpolation. Then smooth the data by Gauss-smooth function. The new $V/I$ maps are shown in Figure \ref{cmap}(b) and (d). It can be seen that the discontinuity has been eliminated. The scatter plots of $|V/I|$ \textit{vs} $I/I_{m}$ are shown in Figure \ref{scatter}(c) and (f). The relationship is approximate linear.

\section{Comparison Between SMFT and HMI data}

The SMFT longitudinal magnetic field $B_{L}^{SMFT}$ can be re-calibrated from equation \ref{eq3}:
\begin{equation}  \label{eq3}
\begin{array}{l}
 B_{L}^{SMFT} =\left\{ \begin{array}{l}
 C_{L}{V \over I}^s \ \ , \ \ \ \ \ \ {I \over I_{m}} < I_{0} \\
 C_{L}{V \over I}  \ \ , \ \ \ \ \ \ \  {I \over I_{m}}  \geq  I_{0}  \\
 \end{array} \right.
 \end{array}
\end{equation}
where $C_{L}$ is calibration coefficient. We adopt 8381 G as proposed by \cite{Su+Zhang+2004}.

\begin{figure}
   \centerline{\includegraphics[width=14cm,clip=]{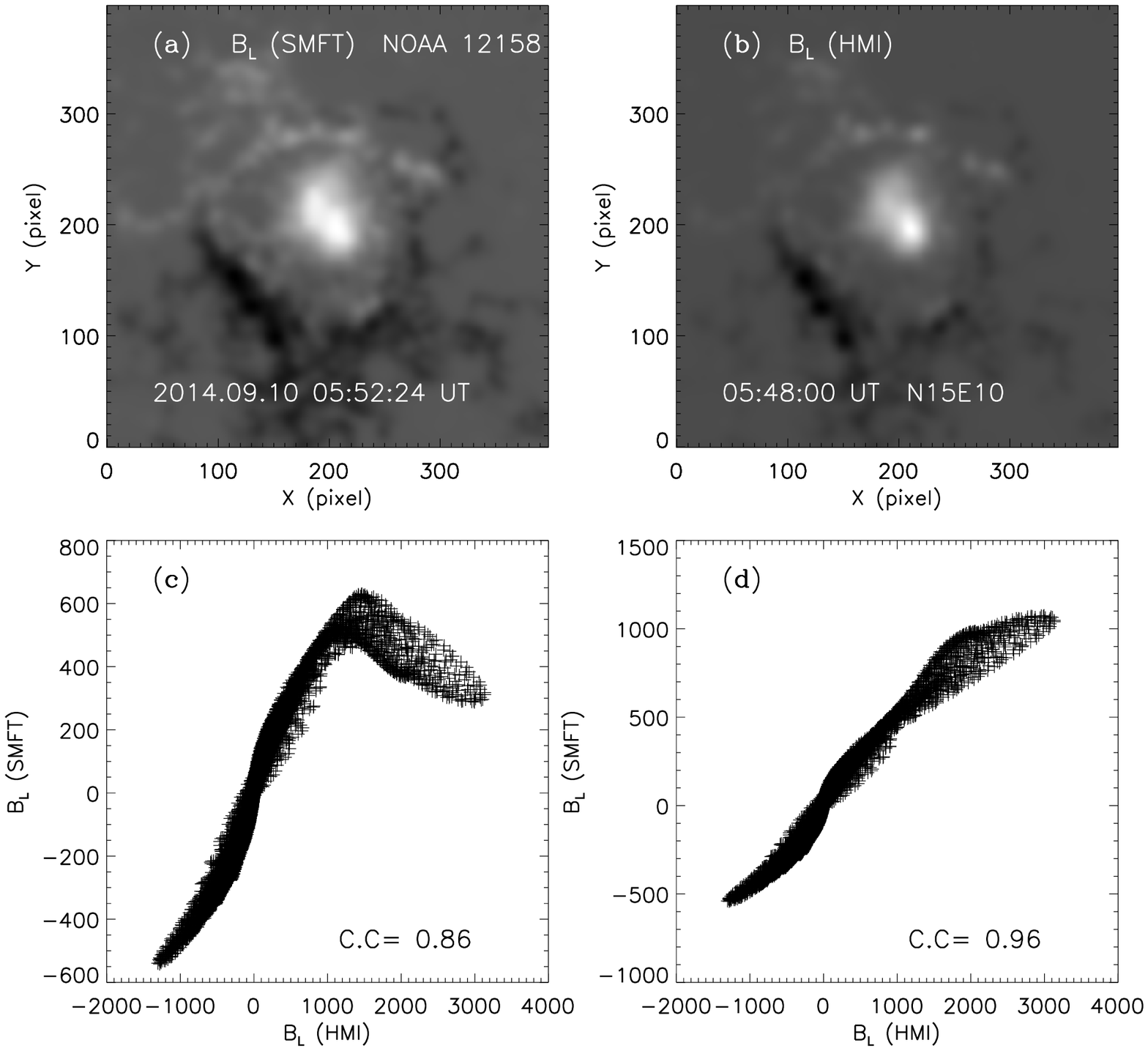}}
   \caption{Panels (a) and (b) are \textbf{$B_{L}$ maps of} NOAA 12158 observed by SMFT and HMI\textbf{, respectively}. \textbf{Panels (c) and (d) present the  scatter plots of $B_{L}$ observed by SMFT and HMI. The data taken by SMFT have saturation in (c) and no saturation in (d).}  C.C is the linear correlation coefficient.} \label{ccb1}
\end{figure}

The comparison of longitudinal magnetic field observed by SMFT and HMI for NOAA12158 is shown in Figure \ref{ccb1}. It can be seen that the distribution of $B_{L}^{SMFT}$ and HMI longitudinal magnetic field $B_{L}^{HMI}$  is very similar (Figure \ref{ccb1}(a) and (b)). The scatter plots of $B_{L}^{SMFT}$ and $B_{L}^{HMI}$ before and after correcting saturation effect are shown in Figure \ref{ccb1}(c) and (d), repectively. The $B_{L}^{SMFT}$ starts to decrease when $B_{L}^{HMI}$ is larger than 1300 G before correcting saturation effect. The linear correlation coefficient is 0.86. After correcting saturation effect in $B_{L}^{SMFT}$, the relationship of $B_{L}^{SMFT}$ and $B_{L}^{HMI}$ is closer to linear. The linear correlation coefficient increases to 0.96. Such good correlation indicates that the proposed correction method for saturation in $B_{L}^{SMFT}$ is effective for this active region.

\begin{figure}
   \centerline{\includegraphics[width=14cm,clip=]{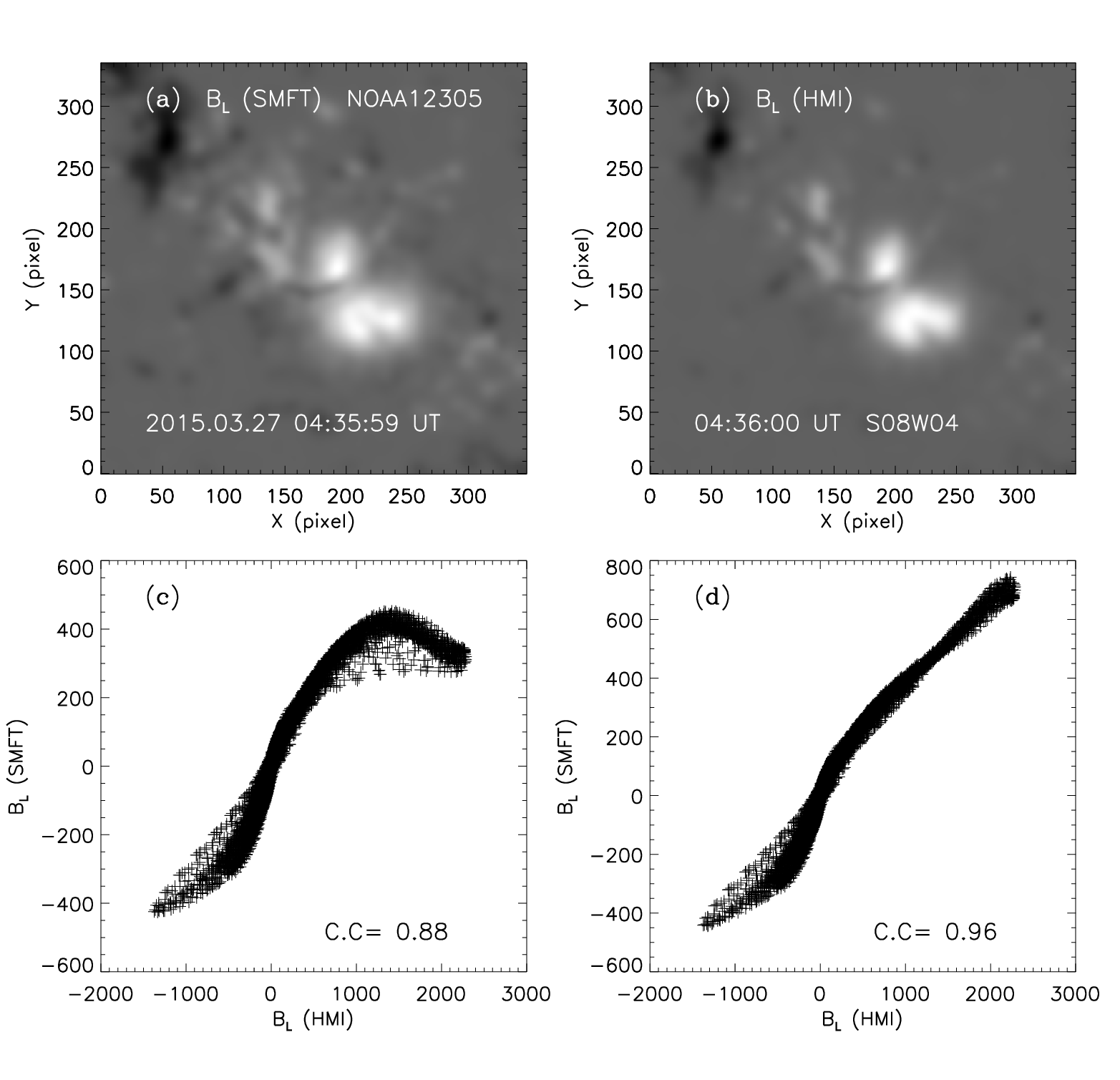}}
   \caption{Similar as Figure \ref{ccb1}, but for NOAA12305.} \label{ccb2}
\end{figure}

Figure \ref{ccb2} shows the comparison of $B_{L}^{SMFT}$ and $B_{L}^{HMI}$ for NOAA12305. It is also found that the $B_{L}^{SMFT}$ starts to decrease when $B_{L}^{HMI}$ is larger than 1300 G before correcting saturation effect. The linear correlation coefficient is 0.88. After correcting saturation effect in $B_{L}^{SMFT}$, a good correlation between $B_{L}^{SMFT}$ and $B_{L}^{HMI}$ is found. The linear correlation coefficient increases to 0.96, which indicates that the proposed correction method for saturation in $B_{L}^{SMFT}$ is also effective for  activity region includes multi-sunspots.

We performed such pixel by pixel comparison for 9 ARs and listed the correlation coefficients in Table \ref{tab2}. The correlation of $B_{L}^{SMFT}$ and $B_{L}^{HMI}$ is much better after eliminating magnetic saturation in B$_{L}^{SMFT}$. So the detection and correction algorithms can be used to re-calibrate the longitudinal magnetic field in strong field region observed by SMFT .

\begin{table}
\bc
\caption[]{Correlation coefficient of $B_{L}$ for 9 ARs observed by SMFT and HMI\label{tab2}}
 \begin{tabular}{ccccccccccc}
  \hline\noalign{\smallskip}
NOAA& Date& Position&C.C (before)&C.C (after)\\
  \hline\noalign{\smallskip}
11658&2013.11.19&S11W10&0.79&0.94\\
11899&2013.11.16&N06W03&0.84&0.93\\
11960&2014.01.25&S14E01&0.89&0.95\\
12027&2014.06.07&N13W01&0.85&0.95\\
12055&2014.05.12&N10W02&0.79&0.93\\
12149&2014.08.27&N10E11&0.88&0.94\\
12158&2014.09.10&N15E10&0.86&0.96\\
12305&2015.03.27&S08W04&0.88&0.96\\
12325&2015.04.19&N04E02&0.88&0.94\\
  \noalign{\smallskip}\hline
\end{tabular}
\ec
\tablecomments{0.7\textwidth}{C.C (before) and C.C (after) represent the linear \textbf{correlation} coefficient before and after correcting magnetic saturation in B$_{L}^{SMFT}$, respectively.}
\end{table}

\section{Testing for large sample}

The algorithm was proved to be completely effective by comparing the results with HMI data for individual active region. To check the applicability of the algorithm for large sample, we \textbf{tested it with} 175 longitudinal magnetograms of 175 ARs observed in 2013 by SMFT. The magnetic saturation generally occurs in strong field region. Considering this actual situation, we set the following restrictions:\\
(1) Only pixels where $|V/I|$ $>$ 0.02 and $I/I_{m}$ $\leq$ 0.7 are used for second-order polynomial fitting.\\
(2) To ensure the rationality of the fitting result, we set $I_{min}$ $<$ $I_{0}$ $<$ 0.7. $I_{min}$ is the minimum value of $I/I_{m}$. If considering 1$\sigma$ error range, we can set $I_{min}$ $<$ $I_{-\sigma}$ $<$  $I_{0}$ $<$ $I_{+\sigma}$ $<$ 0.75.

42 ARs were detected with magnetic saturation. By \textbf{manual testing}, the detected ARs are all correct. Only 1 AR with magnetic saturation was not detected by our method. So the accurate rate of detection is $\sim$99.4\%. If we adjust $I/I_{m}$ range, the above undetected AR can also be detected.  It is found that magnetograms of 5 ARs  \textbf{(the total is 42)} were wrong after correcting saturation effect, which indicates that the accurate rate of correction is $\sim$ 88\%. These 5 ARs are \textbf{either} relatively small or with projection effect. The projection effect is complex, and we \textbf{need to do a} further analysis of \textbf{its influence on saturation.}

\section{Conclusions and Discussions}
\begin{figure}
   \centerline{\includegraphics[width=14cm,clip=]{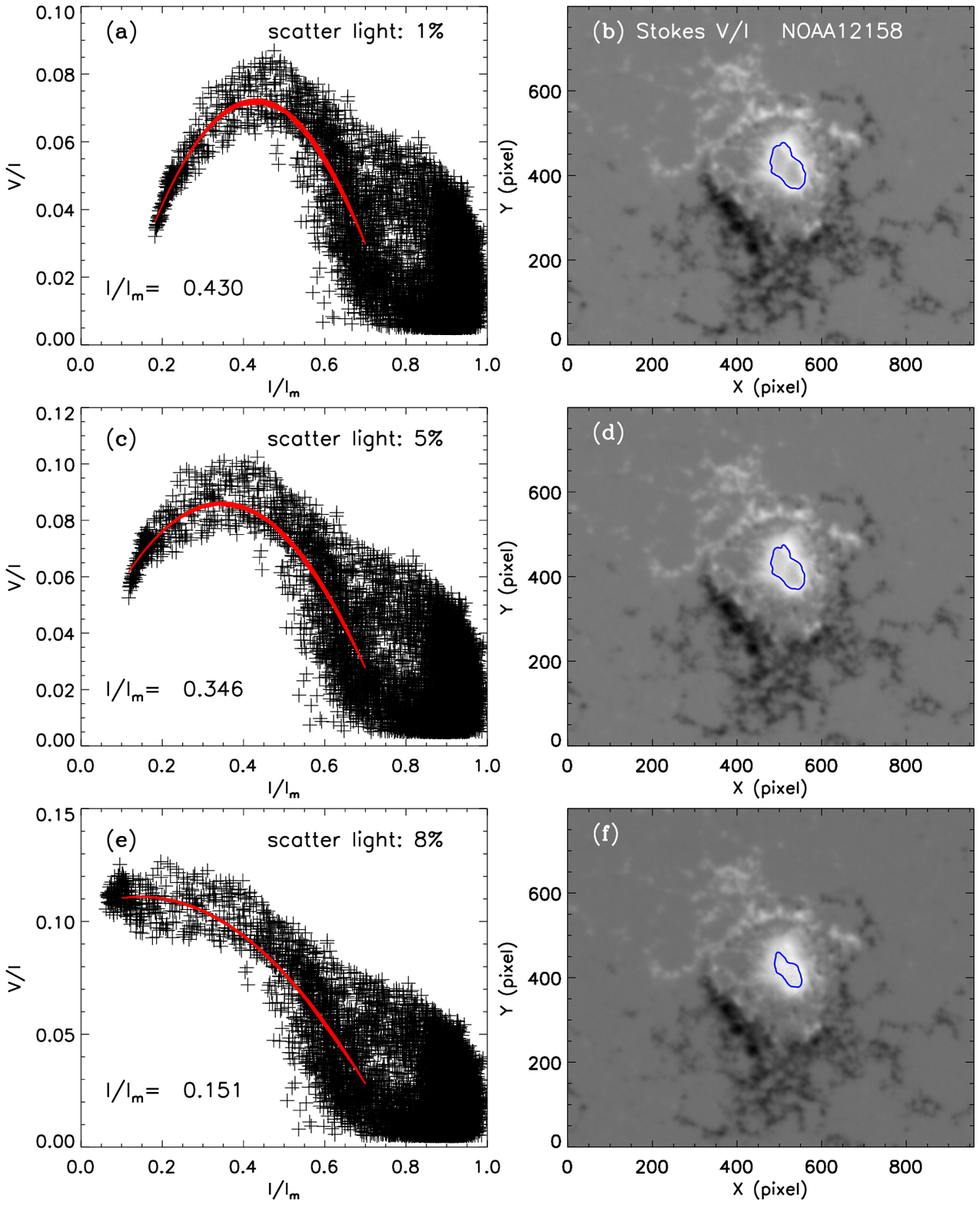}}
   \caption{\textbf{Scatter plots of $|V/I|$ \textit{vs} $I/I_{m}$ and $V/I$ maps. (a) and (b): $I_{s}=I_{c}\times1\%$. (c) and (d): $I_{s}=I_{c}\times5\%$. (e) and (f): $I_{s}=I_{c}\times8\%$. In panels (a), (c) and (e), the red line is the second-order polynomial fit and the marked $I/I_{m}$ corresponds to  apex. The blue contours in $V/I$ maps represent the $I/I_{m}$ apexes  marked in panels (a), (c) and (e).}} \label{slight}
\end{figure}

We developed an automatic detection and correction algorithms for saturation in longitudinal magnetic field observed by SMFT based on the relationship between Stokes $V/I$ and $I$. It works well found in comparison with HMI data in case and sample study. The correlation of longitudinal magnetic fields between SMFT and HMI increased significantly after \textbf{correcting for} saturation effect. The accurate rate of detection and correction is $\sim$99.4\% and $\sim$ 88\% respectively. There are total 43 out of 175 ARs with saturation effect. \textbf{It} means 75.4\% ARs don't need to correct saturation effect.

\textbf{We didn't correct the scatter light when built the $I$-$V/I$ relationship. The measured polarization signals are contaminated by scatter light ($I_{s}$).  E.g., if we consider the scatter light, the equation \ref{eq1} will be written as follow: }
\begin{equation}  \label{eq4}
\begin{aligned}
& {{V} \over {I}} ={{V_{l}-V_{r}}\over {V_{l}+V_{r}-2I_{s}}} \\
& I=V_{l}+V_{r}-2I_{s} \\
\end{aligned}
\end{equation}
\textbf{Generally, $I_{s}$ is determined at the solar limb. Here, we estimate $I_{s}$ using the intensity in quiet sun (a certain percent of $I_{c}$). We took NOAA 12158 as an example to estimate the effect of the scatter light on the method. The result is shown in Figure \ref{slight}. $V/I$ in the umbrae increasing with larger scatter light is subtracted from the observed data. When the contamination level is lower than 8\% (Figure \ref{slight}(a)-(d)), the Stokes $V/I$ \textit{vs} $I/I_{m}$ curves are very similar and the areas of saturation are almost the same although the threshold of saturation is different. It may be due to  the normalized $I/I_{m}$ being  used. The saturation area decreases and the $V/I$ \textit{vs} $I/I_{m}$ curve is close to linear when the contamination level is around 8\% (Figure \ref{slight}(e) and (f)), which shows that the measured polarized signals are likely affected more serious by scatter light than magnetic saturation. The above estimations indicate that the proposed method will not be affected by scatter light when the contamination level is lower, and the scatter light can be corrected as a magnetic saturation effect.}

One advantage of this method is that it can calculate the threshold of saturation and correct it automatically.  Therefore, this method can be used for the routine longitudinal field observations. Another advantage is that the used data \textbf{acquired by} one instrument which avoids a systematic error caused by cross-comparison. Especially, it can be used to correct the saturation effect in longitudinal magnetic fields in past 30 years taken by SMFT . This method can be used for FMG. The disadvantage of these method is that the correction for saturation is not very accurate when the active regions are far from disk center. This may caused by the projection effect. We will improve the method by considering the projection effect in future.


\begin{acknowledgements}
This work is supported by National Natural Science Foundation of China (NSFC)
under No. 11703042, 11911530089, U1731241, 11773038, 11427901, 11427803, 11673033, U1831107, 11873062, the Strategic Priority Research Program on Space Science, the Chinese Academy of Sciences under No. XDA15320302, XDA15052200, XDA15320102, the 13th Five-year Informatization Plan of Chinese Academy of Sciences (Grant No. XXH13505-04). We acknowledge the use of data of SMFT/HSOS and HMI/SDO.
\end{acknowledgements}

\label{lastpage}

\end{document}